\documentclass{edbk}
\usepackage{edbkps}
\usepackage{epsfig}
\let\footnote\savefootnote
\let\footnoterule\savefootnoterule 
\normallatexbib

\begin{document}

\articletitle{Entropy saturation and the Brinkmann-Rice transition in a
random-tiling model}

\author{D. K. Sunko}
\affil{Department of Physics, Faculty of Science, Bijeni{\v c}ka 32,
HR-10000 Zagreb, Croatia.}

\begin{abstract}
The parameter regime in which a Brinkmann-Rice (BR) transition appears near
half-filling is investigated for a model of of one kind of electrons
traversing  a plane randomly tiled with CuO$_4$ molecules, simulating the
copper-oxide planes of high-T$_c$ superconductors. As the hole doping is
increased, the BR transition evolves continuously into a state characterized
by Kauzmann-like plateaus in the entropy vs. temperature curves.  Despite
clear analogies with the glass transition, these are equilibrium properties
of the model. This is because the spin interactions, responsible for ordering
in real space, are not included.
\end{abstract}

\paragraph{Introduction}
One difficulty in describing the metal-insulator transition, caused by strong
electron repulsion, is that there is no obvious classical solution in the
limit of infinite repulsion. This is in contrast to `strong-coupling'
problems with attractive interactions, which are difficult when the
coupling is competitive with the hopping scale, but simplify when it
becomes much greater.

Like some other approaches, saddle-point slave-boson~\cite{Kotliar88} and
dynamical mean-field theory~\cite{Georges96}, the random-tiling (RT)
model~\cite{Sunko96} tries to get around this problem by simply postulating a
`heavy mode'. The RT heavy mode is essentially the one proposed by
Gutzwiller: electrons of one spin see those of the other as static. This is
implemented as a Falicov-Kimball limit of the three-band Emery model, in
which only up-electrons move, by projected hopping. There are no spin
interactions in the model. The parameters are the hopping overlap $t$ and
copper-oxygen splitting $\Delta_{pd}$.

\begin{figure}[htb]
\center{\epsfig{file=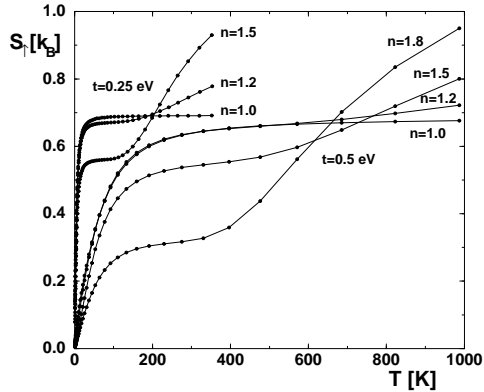,height=60mm}}
\caption{Entropy of mobile spins, for
$t=0.25$~eV (left three curves), and $t=0.5$~eV (right four curves).
Curves are marked by the corresponding concentration.
Saturation plateaus are clearly visible. Here $\Delta_{pd}=3$~eV,
$U=10$~eV, and $n=2n_\uparrow=2n_\downarrow$.}
\label{fig}
\end{figure}

\paragraph{Brinkmann-Rice regime} Here I concentrate on the limit
$t\ll\Delta_{pd}$, which was identified in Ref.~\cite{Sunko96} as having a
crossover of Brinkmann-Rice~\cite{Brinkmann70} type. The novelty is in what
happens when one moves away from half-filling. In Fig.~(\ref{fig}), I show
the entropy of the mobile spins. There appear saturation plateaus,
reminiscent of Kauzmann plateaus~\cite{Kauzmann48} in vitreous liquids. One
of the curves even exhibits a `Kauzmann paradox', extrapolation from the
high-temperature part yielding a negative entropy at low temperatures.
Entropy saturation due to `freezing' of kinetic motion was also observed in
$^3$He, and even modelled by the Hubbard model~\cite{Seiler86}.

The value of the entropy at the plateaus corresponds to complete static
disorder of the tiles, so it is the maximum one-band entropy, the rise beyond
being due to the oxygen degrees of freedom. These play a role analogous to
classical translation in the case of $^3$He. Once the rise occurs, the oxygen
gas rapidly becomes Maxwellian (the chemical potential moves out of the
effective band). The plateaus signal `pseudogaps' in the \emph{system}
density of states. The single-particle model on which the problem is mapped
reflects this by a very narrow effective band for the parameters in question,
so the Maxwellian regime corresponds to the temperature becoming comparable
to this effective band-width. The input parameters, as shown, are still on
the usual `electron' scale, indicating strong model renormalization when
$t/\Delta_{pd}<1/5$. The effective \emph{one-particle} density of states
shows no pseudogaps.

For realistically large values of the hopping overlap, \emph{e.g.} $t=1$~eV
for $\Delta_{pd}=3$~eV, the entropy has a fairly smooth `metallic' rise at
finite fillings, and there is no Brinkmann-Rice regime near half-filling,
either.

\paragraph{Discussion} The Brinkmann-Rice `transition' in the RT model is the
end-point of a continuous range of glass-like crossovers, tuned by doping. 
The model interpolation, between quantum order at low temperatures and
classical disorder at high ones, is not smooth in this parameter regime,
despite the input $t/\Delta_{pd}$ being fairly large, though smaller
than suggested by experiment.

The expression `glass-like' is used only conditionally, to describe the shape
of the entropy curves. In the absence of spin interactions, there is no
spatially ordered state of lower free energy, even at half-filling. In fact,
the RT Mott-Hubbard transition occurs slightly below half-filling,
immediately producing a less ordered state~\cite{Sunko00}. This tendency of
charge correlations to disorder the system, presumably counteracting spin
correlations on the hole-doped side, is one of the more interesting aspects
of the RT model.

Experimentally, a pseudogap can be noticed in the temperature dependence of
the entropy of LSCO~\cite{Loram96}. For underdoped systems, the curve is
first flat, then rises, so that extrapolation from the high-temperature part
indicates a disappearance of states at the Fermi level. However, the
parameter regime needed to obtain the observed \emph{values} of the entropy
in the RT model is the above-mentioned realistic one, $t/\Delta_{pd}\sim
1/3$, where the entropy curves are smooth, not like the ones shown here. This
discrepancy of qualitative and quantitative fits is probably due to the
overly simplistic Falicov-Kimball limit.

\paragraph{Acknowledgements} Conversations with S.~Bari\v{s}i\'{c} and
E.~Tuti\v{s}, and one with T.M.~Rice, are gratefully acknowledged. This work
was supported by the Croatian Government under Project~$119\,204$.

\end{document}